\documentclass[aps,twocolumn,showpacs]{revtex4}
\usepackage{psfig}
\usepackage{feynmp}

\newcommand{\be}{\begin{equation}}
\newcommand{\ee}{\end{equation}}
\newcommand{\ba}{\begin{eqnarray}}
\newcommand{\ea}{\end{eqnarray}}

\newcommand{\ep}{\epsilon}

\begin{document}

\title{
%
%
\[ \vspace{-2cm} \]
\noindent\hfill\hbox{\rm  } \vskip 1pt
\noindent\hfill\hbox{\rm Alberta Thy 18-01} \vskip 1pt
\noindent\hfill\hbox{\rm SLAC-PUB-9100} \vskip 1pt
\noindent\hfill\hbox{\rm hep-ph/0112267} \vskip 10pt
%
%
Expansion of bound-state energies in powers of $m/M$ and $(1-m/M)$
}

\author{Ian Blokland and Andrzej Czarnecki}
\affiliation{
Department of Physics, University of Alberta\\
Edmonton, AB\ \  T6G 2J1, Canada\\
E-mail: blokland@phys.ualberta.ca,  czar@phys.ualberta.ca}

\author{Kirill Melnikov}
\affiliation{Stanford Linear Accelerator Center\\
Stanford University, Stanford, CA 94309\\
E-mail: melnikov@slac.stanford.edu}

\begin{abstract}
Elaborating on a previous letter~\cite{Czarnecki:2000fv},
we use a new approach to compute energy levels of a
non-relativistic bound-state of two constituents, with masses $m$ and
$M$, by systematic expansions --- one in powers of $m/M$ and another in
powers of $(1-m/M)$.  Technical aspects of the calculations are described in detail.
Theoretical predictions are given for
${\mathcal{O}}(\alpha(Z\alpha)^5)$ radiative recoil and ${\mathcal{O}}((Z\alpha)^6)$ pure recoil corrections to the average energy shift and hyperfine splitting
relevant for hydrogen, muonic hydrogen, and muonium.
\end{abstract}

\maketitle

\section{Introduction}

Precision studies, both theoretical and experimental, of non-relativistic QED bound-states have historically provided a wealth of information about fundamental physical parameters
such as the fine structure constant and the masses of the electron, muon
and proton~\cite{Mohr99}.  In order to keep pace with the ever-improving precision of current experiments, increasingly intricate theoretical calculations must be performed.  Fortunately, the emergence of new calculational techniques has brought many such calculations within reach. 

In a previous letter~\cite{Czarnecki:2000fv}, we introduced a practical algorithm which allows a calculation of
the bound-state energy levels in a given order of perturbation theory (in $\alpha$ and $Z\alpha$) as an expansion in powers and logarithms of $m/M$ with an arbitrary precision.  The opposite situation,
the calculation of the energy levels to all orders in $\alpha$ but in a fixed order in the ratio $m/M$, has been studied in the literature \cite{braun,shabaev,Yelkhovsky:1994ag,PG95}.  We illustrated the method by calculating the ${\mathcal{O}}(\alpha(Z\alpha)^5)$ radiative recoil corrections to the average energy shift and hyperfine splitting of a non-relativistic QED bound-state.  This analytic result enabled us to resolve a discrepancy between two previous calculations of ${\mathcal{O}}(\alpha(Z\alpha)^5m^2/M)$ corrections to the average energy shift, thereby removing a major source of theoretical uncertainty in the isotope shift (i.e. the difference between the $2S$ to $1S$ transition energies in deuterium and hydrogen).

In this paper we extend our previous results in several ways.  For instance, we illustrate how the energy shifts can also be expanded as a series in $(1-m/M)$.  Beyond its applicability to situations where $m \sim M$, this expansion method provides a useful cross-check on the comparably more difficult method of expanding in $m/M$.  We also examine the convergence properties of these expansions in order to ascertain some general guidelines about the accuracy of the truncated series that necessarily arise when using these expansions.  In addition, we present a calculation of the pure recoil corrections using an expansion in $m/M$.

The rest of this paper is organized as follows.  In Section II we discuss the framework of the calculation and our expansion method in general terms.  Section III contains a detailed technical description of the calculation.  In Section IV we present and discuss the results of our calculations for the ${\mathcal{O}}(\alpha(Z\alpha)^5)$ radiative recoil and ${\mathcal{O}}((Z\alpha)^6)$ pure recoil corrections to the average energy shift and hyperfine splitting of a generic QED bound-state.  Our results are summarized in Section V.  Finally, the Appendix illustrates the techniques with which a class of loop integrals --- the so-called eikonal integrals --- can be evaluated.

\section{Framework of the Calculation}

Non-relativistic bound-states can be conveniently described by an effective field theory which exploits a separation in the energy scales $m$, $mv$, and $mv^2$~\cite{Caswell:1986ui}.
This approach is facilitated further by the use of dimensional regularization~\cite{Pineda:1998bj,Pineda:1998kn,Czarnecki:1998zv,Czarnecki:1999mw}.  The basic idea is that the Coulomb potential supplies the dominant interaction and all other interactions provide corrections which can be evaluated using the familiar time-independent perturbation theory of quantum mechanics.  These corrections can be divided into two classes.  

The first class is that of the so-called ``soft'' contributions, governed by long-range potential terms.  These contributions can be evaluated for arbitrary masses of the constituent particles because the essential soft dynamics of a non-relativistic bound-state, as described by the Schr\"odinger equation, are characterized by the reduced mass of the system rather than the individual masses of the consituents.  As a result, once the soft contributions are obtained in the equal mass case~\cite{Czarnecki:1998zv,Czarnecki:1999mw}, the more general mass case follows easily.

The ``hard'' contributions make up the second class, and these contributions lead to the interactions that can be characterized by $\delta(r)$ potential terms.  Such terms result from the relativistic region of loop-momentum integrals, and they are usually obtained as Taylor expansions of scattering amplitudes in terms of the spatial momentum components of the external particles, which are taken to be on-shell.  At lowest order in $\alpha$, the hard diagrams should be evaluated exactly at threshold, whereby the constituents have zero relative velocity.  This implies that the relevant loop-momentum integrals depend on only two scales, $m$ and $M$.  The hard contributions have a much more complicated dependence on the mass scales than the soft contributions do, and this is why we will expand the hard scattering diagrams in powers of either $m/M$ or $(1-m/M)$.  By expanding the \emph{integrands}, we are left with only homogeneous, one-scale integrals to evaluate, and this consitutes a substantial simplification.  This method lends itself to automation so that many terms of the expansion can be obtained, with the only limitation being set by the available computing power.  High-performance symbolic algebra software is of great
help in such computations (we use FORM \cite{form3}).  

Our $m/M$ expansion method is motivated by a procedure in which Feynman diagrams are expanded in large masses and momenta~\cite{Chetyrkin91,Tkachev:1994gz,Smirnov:1995tg}.  Although this procedure was originally expressed in a different way, it can be
reformulated more practically using the notion of momentum regions.
The algorithm, which is applied directly to the loop integrals, consists of five steps~\cite{Beneke:1998zp}.  First, identify the large and small external scales in the integrals.  Second, divide the integration volume into regions so that the momentum flow through any of the internal lines is of the order of one of the external scales.  More specifically, the statement $k \sim M$ asserts that $k^2 > m^2$ in Euclidean space.  Third, perform Taylor expansions within every region for any individual denominator factors (propagators) where the terms within these factors depend differently on the external scales.  Fourth, integrate the expanded integrands from every region over the initial integration volume --- in other words, ignoring the constraints that identify the regions.  Finally, add the contributions arising from the individual regions in order to obtain the final result.  The fourth step of this algorithm requires further explanation, since by ignoring the constraints on the individual regions, it may appear as if contributions to the total integral are counted more than once.  This does not happen because the extra contributions to the total integral that are introduced by removing the constraints on individual regions can be expressed as scale-less integrals, and scale-less integrals vanish in dimensional regularization.  This implies that the integrals from the various regions are different analytic functions of the parameters of the problem.  In the next section, we will demonstrate this algorithm in detail.

\section{Procedure}

\begin{figure}[ht]
  \begin{center}
  \begin{fmffile}{rr1}
  \begin{fmfgraph*}(140,40)
    \fmfleft{i1,i2}
    \fmfright{o1,o2}
    \fmf{electron}{i2,v1}
    \fmf{plain}{v1,v2,v3,v4,v5,v6}
    \fmf{electron}{v6,o2}
    \fmf{electron,width=2}{o1,v8}
    \fmf{plain,width=2}{v8,v7}
    \fmf{electron,width=2}{v7,i1}	
    \fmffreeze
    \fmf{photon,left=0.6}{v2,v5}
    \fmf{photon}{v1,v7}
    \fmf{photon}{v6,v8}           	            
  \end{fmfgraph*}
  \end{fmffile}
  \vspace{10mm}
  \begin{fmffile}{rr2}
  \begin{fmfgraph*}(140,40)
    \fmfleft{i1,i2}
    \fmfright{o1,o2}
    \fmf{electron}{i2,v1}
    \fmf{plain}{v1,v2,v3,v4}
    \fmf{electron}{v4,o2}
    \fmf{electron,width=2}{o1,v6}
    \fmf{plain,width=2}{v6,v5}
    \fmf{electron,width=2}{v5,i1}	
    \fmffreeze
    \fmf{photon,left=0.5}{v1,v4}
    \fmf{photon}{v2,v5}
    \fmf{photon}{v3,v6}           	            
  \end{fmfgraph*}
  \end{fmffile}
  \vspace{10mm}
  \begin{fmffile}{rr3}
  \begin{fmfgraph*}(140,40)
    \fmfleft{i1,i2}
    \fmfright{o1,o2}
    \fmf{electron}{i2,v1}
    \fmf{plain}{v1,v2,v3,v4}
    \fmf{electron}{v4,o2}
    \fmf{electron,width=2}{o1,v8}
    \fmf{plain,width=2}{v5,v6,v7,v8}
    \fmf{electron,width=2}{v5,i1}	
    \fmffreeze
    \fmf{photon,left=0.5}{v1,v3}
    \fmf{photon}{v2,v6}
    \fmf{photon}{v4,v8}
  \end{fmfgraph*}
  \end{fmffile}
  \end{center}
\caption{The forward-scattering radiative-recoil diagrams. The bold
line represents the heavy constituent of the bound-state (e.g. a proton
if we consider hydrogen) and the thin line --- the light one (an
electron). Diagrams with the crossed photons in the $t$-channel are
not displayed.}
\label{fig1}  
\end{figure}
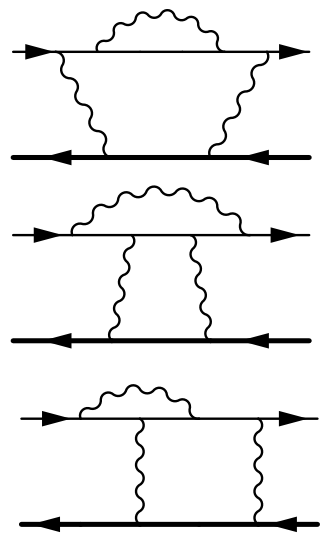

In the previous letter~\cite{Czarnecki:2000fv}, we outlined the expansion procedure for the radiative recoil diagrams.  These diagrams are shown in Fig.~\ref{fig1}.

\vspace{5mm}

\begin{figure}[ht]
  \begin{center}
  \begin{fmffile}{r1}
  \begin{fmfgraph*}(140,40)
    \fmfleft{i1,i2}
    \fmfright{o1,o2}
    \fmf{electron}{i2,v1}
    \fmf{plain}{v1,v2,v3}
    \fmf{electron}{v3,o2}
    \fmf{electron,width=2}{o1,v6}
    \fmf{plain,width=2}{v6,v5,v4}
    \fmf{electron,width=2}{v4,i1}	
    \fmffreeze
    \fmf{photon}{v1,v4}
    \fmf{photon}{v2,v5}
    \fmf{photon}{v3,v6}           	            
  \end{fmfgraph*}
  \end{fmffile}
  \vspace{10mm}
  \begin{fmffile}{r2}
  \begin{fmfgraph*}(140,40)
    \fmfleft{i1,i2}
    \fmfright{o1,o2}
    \fmf{electron}{i2,v1}
    \fmf{plain}{v1,v2,v3}
    \fmf{electron}{v3,o2}
    \fmf{electron,width=2}{o1,v6}
    \fmf{plain,width=2}{v6,v5,v4}
    \fmf{electron,width=2}{v4,i1}	
    \fmffreeze
    \fmf{photon}{v1,v6}
    \fmf{photon}{v2,v5}
    \fmf{photon}{v3,v4}           	            
  \end{fmfgraph*}
  \end{fmffile}
  \vspace{10mm}
  \begin{fmffile}{r3}
  \begin{fmfgraph*}(140,40)
    \fmfleft{i1,i2}
    \fmfright{o1,o2}
    \fmf{electron}{i2,v1}
    \fmf{plain}{v1,v2,v3}
    \fmf{electron}{v3,o2}
    \fmf{electron,width=2}{o1,v6}
    \fmf{plain,width=2}{v6,v5,v4}
    \fmf{electron,width=2}{v4,i1}	
    \fmffreeze
    \fmf{photon}{v1,v6}
    \fmf{photon}{v2,v4}
    \fmf{photon}{v3,v5}               	            
  \end{fmfgraph*}
  \end{fmffile}
  \vspace{10mm}
  \begin{fmffile}{r4}
  \begin{fmfgraph*}(140,40)
    \fmfleft{i1,i2}
    \fmfright{o1,o2}
    \fmf{electron}{i2,v1}
    \fmf{plain}{v1,v2,v3}
    \fmf{electron}{v3,o2}
    \fmf{electron,width=2}{o1,v6}
    \fmf{plain,width=2}{v6,v5,v4}
    \fmf{electron,width=2}{v4,i1}	
    \fmffreeze
    \fmf{photon}{v1,v4}
    \fmf{photon}{v2,v6}
    \fmf{photon}{v3,v5}
    \fmflabel{$p_{1}$}{i2}
    \fmflabel{$-p_{2}$}{i1}
    \fmflabel{$p_{1}$}{o2}
    \fmflabel{$-p_{2}$}{o1}               	            
  \end{fmfgraph*}
  \end{fmffile}  
  \vspace{5mm}    
  \end{center}
\caption{The forward-scattering pure recoil diagrams.}
\label{fig2}  
\end{figure}
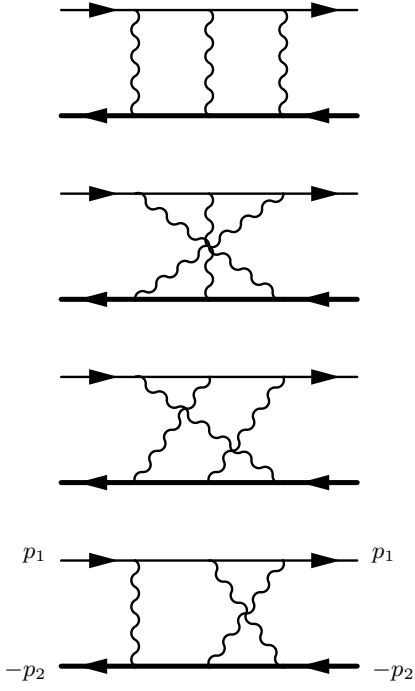

We will now describe the expansion procedure using the pure recoil diagrams, since a few additional complications arise.  These diagrams are shown in Fig.~\ref{fig2}.

To illustrate the method we focus on the last diagram
in Fig.~\ref{fig2} and consider the following scalar
integral:
\ba
&&\int
\frac {[{\rm d}^D k_1] [{\rm d}^D k_2]}
{(k_1^2) (k_2^2) (k_1+k_2)^2 \left[(k_1 + k_2)^2 + 2(k_1 + k_2 )p_1 \right] }
\label{exint}
\\
&& \times \frac {1}{ (k_1^2 + 2k_1 p_1)
(k_1^2 - 2k_1 p_2 + i \delta) (k_2^2 + 2k_2 p_2 + i \delta)} \ . \nonumber
\ea
Here $[{\rm d}^D k]$ stands for ${\rm d}^D k/(2\pi)^D$, $p_1 \equiv mQ$,
$p_2 \equiv MQ$, where $Q=(1,0,0,0)$ is the time-like unit
vector.  Only the relevant infinitesimal imaginary parts of the propagators have been displayed.  We are going to illustrate the 
expansion of the integral in Eq.~(\ref{exint}) in powers of $m/M$ following 
the five steps outlined above.

There are five momentum regions to be considered. In the first one
all the momenta are of the order of the large mass $M$. In this
case one can expand the electron propagators in $mQk_i$. The resulting
integrals are all of the form:
\be
\label{region1}
\int \frac {[{\rm d}^D k_1] [{\rm d}^D k_2]}
{{(k_1^2)}^{a_1} {(k_2^2)}^{a_2} {(k_1+ k_2)}^{2a_3}
(k_1^2 - 2k_1p_2)^{a_4} (k_2^2 + 2k_2p_2)^{a_5}},
\ee
with some integer powers $a_i$. One immediately recognizes that
all these integrals are identical with the general two-loop self-energy
integrals of the particle with mass $M$ for which the general
solution is known \cite{bro91a}.

Next, there is a momentum region where $k_1 \sim M$ and $k_2 \sim m$.  By using $k_1 \sim M$ to expand the electron propagators we obtain Eq.~(\ref{region1}) again.  With $k_2 \sim m$, the $(k_1 + k_2)^2$ and $(k_2^2 + 2k_2p_2)$ factors can also be expanded, so that the only denominator factors which depend on $k_2$ are $(k_2^2)$ and $(2k_2p_2)$.  Since we are working at threshold, we have $p_2 \equiv MQ$, which leads to
\be
\label{eikzero}
\int \frac {[{\rm d}^D k_2]}
{{(k_2^2)}^{\alpha} (2k_2p_2 + i\delta)^{\beta}} = 0 \ ,
\ee 
so that this region provides no contribution to the amplitude for this particular diagram.

The third momentum region has $k_1 \sim M$ and $k_2 \sim M$ but $k_1 + k_2 \sim m$.  After a Taylor expansion in small variables, the integrals in this region factorize into products of two simple one-loop integrals.

In the fourth region, $k_1 \sim m$ and $k_2 \sim M$.  A Taylor expansion in small variables allows the integrals in this region to be factored into one-loop integrals as
\ba
&& \int \frac{[{\rm d}^D k_1]}{(k_1^2)^{a_1}(k_1^2 + 2k_1 p_1)^{a_2} (2k_1 p_1)^{a_3}} \nonumber \\
&& \times \int \frac{[{\rm d}^D k_2]}{(k_2^2)^{a_4}(k_2^2 + 2k_2 p_2)^{a_5}}.
\ea
The $k_2$ integral is a trivial one-loop integral.  The $k_1$ integral can be converted to the same simple form, along with integrals like Eq.~(\ref{eikzero}), by multiplying it by factors of
\be
1 = \frac{(k_1^2 + 2k_1 p_1)}{(k_1^2)} - \frac{(2k_1 p_1)}{(k_1^2)}
\ee
until either $a_2$ or $a_3$ is brought to zero.

The fifth region is characterized by the condition $k_1 \sim k_2 \sim m$.  In this case, the heavy particle propagators can be expanded into \emph{static}, or as we will call them, \emph{eikonal}, propagators.  The integrals in this region are of the form
\ba
\label{region5}
&& \int \frac{[{\rm d}^D k_1][{\rm d}^D k_2]}{(k_1^2)^{a_1} (k_2^2)^{a_2} (k_1 + k_2)^{2a_3} \left[ (k_1 + k_2)^2 + 2(k_1 + k_2)p_1 \right]^{a_4}} \nonumber \\
&& \times \frac{1}{(k_1^2 + 2k_1p_1)^{a_5} (2k_1p_2 - i\delta)^{a_6} (2k_2p_2 + i\delta)^{a_7}} \ .
\ea
Notice how the eikonal propagators arising from $(k^2-2kp+i\delta)$ factors acquire $-i\delta$ pole terms.  Such terms are important in this region and must be carefully accounted for.  The integral in Eq.~(\ref{region5}) can be simplified using the identity
\be
\frac{1}{(k^2+2kp)(2kp)} = \frac{1}{k^2} \left( \frac{1}{(2kp)} - \frac{1}{(k^2+2kp)} \right).
\ee 
Once one of the seven factors in Eq.~(\ref{region5}) has been removed, an identity can be constructed from the observation that the remaining six factors are linearly dependent.  Using such an identity, the integrals in this fifth region can be expressed as one of four types of integrals.  The first type is the two-loop self energy integral of a particle with mass $m$ for which the general solution is known~\cite{bro91a}.  The remaining three types contain eikonal propagators:
\ba
\label{E1}
E_1^{\pm} & = & \int \frac{[{\rm d}^D k_1][{\rm d}^D k_2]}{ (k_1^2)^{a_1} (k_2^2)^{a_2} (k_1 - k_2)^{2a_3}}  \\
&& \times \frac{1}{(k_2^2 + 2k_2p_1)^{a_4} (2k_1p_1 \pm i\delta)^{a_5}} \ ,
\nonumber \\
\label{E2}
E_2^{\pm} & = & \int \frac{[{\rm d}^D k_1][{\rm d}^D k_2]}{ (k_1^2)^{a_1} (k_2^2)^{a_2} (k_1^2 + 2k_1p_1)^{2a_3}} \\
&& \times \frac{1}{\left[(k_1 + k_2)^2 + 2(k_1 + k_2)p_1 \right]^{a_4} (2k_2p_1 \pm i\delta)^{a_5}} \ ,
\nonumber  \\
\label{E3}
E_3^{\pm \pm} & = & \int \frac{[{\rm d}^D k_1][{\rm d}^D k_2]}{ (k_1^2)^{a_1} (k_2^2)^{a_2} (k_1 - k_2)^{2a_3}}  \\
&& \times \frac{1}{(2k_1p_2 \pm i\delta)^{a_4} (2k_2p_2 \pm i\delta)^{a_5}} \ . \nonumber
\ea  
At threshold, the $E_3$ integrals are exactly zero for similar reasons as are needed to establish  Eq.~(\ref{eikzero}).  This leaves us with two new types of integrals, $E_1$ and $E_2$, required for the recoil calculations.  The calculations for radiative recoil diagrams involve $E_2$ but not $E_1$ integrals.  In Appendix A, we outline the procedure by which the $E_1$ and $E_2$ integrals can be evaluated.

We shall now describe a second method of expansion, relevant to the scenario where the two bound-state constituents have similar, but not necessarily equal, masses.  Although this scenario is not realized by any common QED bound-states, this second expansion can provide a useful check on the first expansion method.

The essential idea is to introduce an expansion parameter
\be
y = 1 - \frac{m}{M}
\ee
so that the external momentum of the light particle, $p_{1}$, can be written in terms of $y$ and the external momentum of the heavy particle, $p_{2}$, via
\be
p_{1} = (1-y)p_{2} \ .
\ee 
Then, any massive propagator containing $p_{1}$ can be expanded, as a series in powers of $y$, in terms of the corresponding propagators containing $p_{2}$:
\be
\frac{1}{k^{2}+2kp_{1}} = \sum_{n=0}^{\infty} \frac{(2ykp_{2})^{n}}{(k^{2}+2kp_{2})^{n+1}} \ .
\ee
As a result, the two-scale general scalar integral in Eq.~(\ref{exint}) is expanded, as a series in powers of $y$, in terms of two-loop on-shell self-energy integrals.  As an additional cross-check, we note that in the limit that the two masses are equal, we have $y=0$, so that the positronium results \cite{Czarnecki:1998zv,Czarnecki:1999mw} can be recovered from the leading term of this expansion.    

Another method has been proposed~\cite{Jentschura:2001ep} for dealing with similar problems involving more than one mass/energy/distance scale.  Intermediate parameters are introduced to separate various scales and the calculations are performed in four dimensions.  An advantage of such an approach is that it avoids various complications arising when working in $D$ dimensions.  However, it spoils the homogeneity of integrals and it is not clear whether one can apply integration-by-parts algorithms, which are crucial for larger calculations involving many terms of expansions.

\section{Results and Discussion}

We have applied our algorithms to compute the ${\cal O}(\alpha (Z\alpha)^5)$
radiative recoil corrections to the average energy shift and the hyperfine
splitting of a general QED bound-state composed of two spin-1/2 particles 
with masses $m$ and $M$.  In this case 
the soft contribution is absent and the hard corrections 
shown in Fig.~1 are the only diagrams we have to consider.
We have done the calculation in a general covariant gauge; the 
cancellation of the gauge parameter dependence serves as 
a check of the computation.

For the $S$-wave ground state energy $E$ we define
\be
E = E_{\rm aver} + \left ( \frac {1}{4} - \delta_{J0} \right ) E_{\rm hfs} \ ,
\ee
where $J=0,1$ is the total spin of the two fermions forming the bound-state.

For the hyperfine splitting we obtain
\ba
&& \delta E_{\rm hfs}^{\rm rad~rec} \simeq \frac {8 (Z\alpha)^4 \mu^3}{3mM} 
\alpha (Z\alpha) 
\left \{ \ln 2 - \frac {13}{4}
\right.
\nonumber \\
&& \left.
+ \frac {m}{M} \left ( \frac {15}{4\pi^2} \ln \frac {M}{m}
  + \frac {1}{2}
 + \frac {6 \zeta_3}{\pi^2} + \frac {17}{8 \pi^2} + 3 \ln 2 \right )
\right. \nonumber \\
&& 
 -\left({m\over M}\right)^2 \left ( \frac {3}{2} + 6 \ln 2 \right )
\nonumber \\ &&
+ \left({m\over M}\right)^3 \left (
\frac {61}{12 \pi^2 } \ln^2 \frac {M}{m}
+ \frac {1037}{72 \pi^2 } \ln \frac {M}{m}
\right. 
\nonumber \\
&&\qquad  \qquad  
\left.+  \frac {133}{72}
+ \frac {9\zeta_3}{2\pi^2} + \frac {5521}{288\pi^2} + 3 \ln 2
\right )
\nonumber \\
&&
 - \left({m\over M}\right)^4 \left(   {163\over 48} + 6\ln 2 \right)
\nonumber \\
&&
   + \left({m\over M}\right)^5 \left( 
{331\over 40\pi^2}\ln^2{M\over m}
 + {5761\over 300\pi^2}\ln{M\over m}  +{691\over 240}  
\right.  
\nonumber \\
&&\left. \qquad \qquad
+ {9\zeta_3 \over 2\pi^2} + {206653\over 8000\pi^2} + 3\ln 2
\right)
\nonumber \\
&& \left. - \left( \frac{m}{M} \right)^6
\left( \frac{577}{120} + 6\ln 2 \right) \right\} , 
\label{hfs-r}  
\ea
where $\mu = mM/(m+M)$ is the reduced mass of the bound-state.

For the spin-independent energy shift  we find
\ba
&& \delta E_{\rm aver}^{\rm rad~rec} \simeq \alpha (Z\alpha)^5
 \frac {\mu^3}{m^2} \left \{
  \frac {139}{32} - 2 \ln 2
\right. \nonumber \\
&& \left.
+ \frac {m}{M} \left ( \frac {3}{4} + \frac {6 \zeta_3}{\pi^2}
    - \frac {14}{\pi^2} - 2 \ln 2  \right )
\right.  \nonumber \\
&& 
+ \left({m\over M}\right)^2 \left (  - \frac {127}{32} + 8\ln 2 \right )
\nonumber \\ &&
+\left({m\over M}\right)^3
 \left (  - \frac {8}{3 \pi^2 } \ln^2 \frac {M}{m}
 - \frac {55}{18 \pi^2 } \ln \frac {M}{m}+ \frac {47}{36}
\right. 
\nonumber \\
&& \left . 
\qquad \qquad  -  \frac {3\zeta_3}{\pi^2} - \frac {85}{9\pi^2 }
 - 2 \ln 2 \right )
\nonumber \\
&&
+
\left({m\over M}\right)^4 \left(  - {55\over 24} + 4\ln 2 \right)
\nonumber \\
&&
 + \left({m\over M}\right)^5 \left({37\over 60\pi^2}\ln^2{M\over m}  
+ {29\over 900\pi^2}\ln{M\over m} + {1027\over 360}
\right.
\nonumber \\ &&
 \qquad \qquad \left. - {3\zeta_3\over \pi^2} - {370667\over 36000\pi^2} - 2\ln 2 \right)
\nonumber \\ &&
         + \left({m\over M}\right)^6 \left(  - {67\over 20} + 4\ln 2 \right) 
\nonumber \\
&& + \left( \frac{m}{M} \right)^7 \left( \frac{199}{70\pi^2} \ln^2{M\over m}
- \frac{1759}{7350\pi^2}\ln{M\over m} + \frac{887}{210} \right.
\nonumber \\
&& \qquad \qquad \left. \left. - \frac{3\zeta_3}{\pi^2} - \frac{241491119}{18522000\pi^2} - 2\ln 2 \right) \right\} . 
\label{avg-r} 
\ea

To our knowledge the terms ${\cal O}(m^3/M^3)$ and higher are new for
both $E_{\rm hfs}$ and $E_{\rm aver}$, while the other terms have been
obtained previously~\cite{Eides:2000xc}.  The coefficient of the ${\cal O}(m/M)$
term in Eq.~(\ref{avg-r}) was the subject of some controversy, since
two different numerical results have been reported,
\cite{Bhatt1,Bhatt2,Bhatt3} and \cite{KPLamb}.

Our result for this term,
\ba
&& \alpha (Z\alpha)^5
 \frac {\mu^3}{m^2} \frac {m}{M} \left ( \frac {3}{4}
+ \frac {6 \zeta_3}{\pi^2} - \frac {14}{\pi^2} - 2 \ln 2  \right )
\nonumber \\
&& \simeq -1.32402796~\alpha (Z\alpha)^5
 \frac {\mu^3}{m^2} \frac {m}{M},
\label{e10}
\ea
is in excellent agreement with the numerical result of
Ref.~\cite{KPLamb} where the coefficient 
$-1.324029(2)$ was obtained, and has since been confirmed in an independent analytical calculation~\cite{Eides:2000kj}. 

\begin{figure}[htp]
\psfig{figure=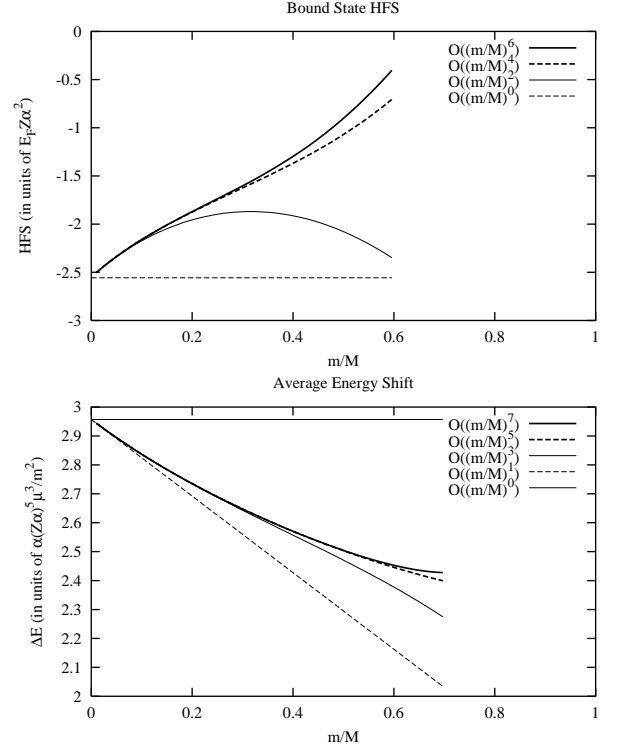,width=90mm}
\vspace*{3mm}
\caption{${\mathcal{O}}(\alpha(Z\alpha)^5)$ radiative recoil contributions to the hyperfine splitting and average energy shift in the $m/M$ expansion.}
\label{r-fig}
\end{figure}

The expansions in $m/M$ of Eqs.~(\ref{hfs-r}) and (\ref{avg-r}) do not yield accurate results for increasing values of $m/M$, even though an untruncated series can be expected to converge on the interval $m/M \in [0,1)$.  The convergence of the terms in the series which we have calculated is depicted in Fig.~\ref{r-fig}.  The upper graph shows the hyperfine splitting calculations as a function of $m/M$ and the lower graph shows the corresponding results for the average energy shift.  In both cases, several curves are plotted, each representing the sum of the first $N$ terms of the expansion, where $N$ is shown in the legend.  The graphs suggest that more terms in the series would be required to obtain reliable values of the hyperfine splitting for $m/M$ larger than about 0.2; our series for the spin-independent energy shift should be reliable for $m/M$ values up to about 0.5.   

The expansion in $y=1-m/M$ aims to address the region where $m/M$ is large.  For the hyperfine splitting we find
\ba
&& \delta E_{\rm hfs}^{\rm rad~rec} \simeq \frac{8(Z\alpha)^4\mu^3}{3mM} \left\{ 
\left( \frac{3\zeta_{3}}{4\pi^2} + \frac{7}{8\pi^2} + 2\ln 2 - \frac{79}{32} \right) \right. \nonumber \\
&& + \left( 1 - \frac{m}{M} \right) 
\left( -\frac{3\zeta_{3}}{4\pi^2} - \frac{21}{8\pi^2} - \frac{3}{4}\ln 2 + \frac{25}{96} \right)
\nonumber \\
&& + \left( 1 - \frac{m}{M} \right)^{2} 
\left( -\frac{19}{24\pi^2} - \frac{3}{8}\ln 2 - \frac{19}{384} \right)
\nonumber \\ 
&& + \left( 1 - \frac{m}{M} \right)^{3} 
\left( -\frac{1}{8\pi^2} - \frac{3}{16}\ln 2 - \frac{55}{576} \right)
\nonumber \\ 
&& + \left( 1 - \frac{m}{M} \right)^{4} 
\left( \frac{57}{400\pi^2} - \frac{3}{32}\ln 2 - \frac{2147}{23040} \right)
\nonumber \\ 
&& + \left( 1 - \frac{m}{M} \right)^{5} 
\left( \frac{11}{40\pi^2} - \frac{3}{64}\ln 2 - \frac{631}{7680} \right)
\nonumber \\
&& + \left( 1 - \frac{m}{M} \right)^{6} 
\left( \frac{41389}{117600\pi^2} - \frac{3}{128}\ln 2 - \frac{46679}{645120} \right)
\nonumber \\ 
&& + \left( 1 - \frac{m}{M} \right)^{7} 
\left( \frac{141709}{352800\pi^2} - \frac{3}{256}\ln 2 - \frac{10561}{161280} \right)
\label{hfs-y}
\\
&& + \left. \left( 1 - \frac{m}{M} \right)^{8} 
\left( \frac{5539481}{12700800\pi^2} - \frac{3}{512}\ln 2 - \frac{157753}{2580480} \right) 
\right\} \ , \nonumber 
\ea
and for the spin-independent energy shift we obtain
\ba
&& \delta E_{\rm aver}^{\rm rad~rec} \simeq \alpha (Z\alpha)^5 \frac {\mu^3}{m^2} \left\{ 
\left( \frac{9\zeta_{3}}{2\pi^2} - \frac{35}{4\pi^2} + \frac{31}{12} \right) \right. \nonumber \\
&& + \left( 1 - \frac{m}{M} \right) 
\left( -\frac{9\zeta_{3}}{2\pi^2} + \frac{39}{2\pi^2} - \frac{7}{2}\ln 2 + \frac{45}{32} \right)
\nonumber \\
&& + \left( 1 - \frac{m}{M} \right)^{2} 
\left( \frac{61}{24\pi^2} + \frac{9}{4}\ln 2 - \frac{119}{72} \right)
\nonumber \\ 
&& + \left( 1 - \frac{m}{M} \right)^{3} 
\left( -\frac{29}{24\pi^2} + \frac{1}{8}\ln 2 + \frac{31}{288} \right)
\nonumber \\ 
&& + \left( 1 - \frac{m}{M} \right)^{4} 
\left( \frac{233}{600\pi^2} + \frac{1}{16}\ln 2 - \frac{571}{11520} \right)
\nonumber \\ 
&& + \left( 1 - \frac{m}{M} \right)^{5} 
\left( \frac{509}{1200\pi^2} + \frac{1}{32}\ln 2 - \frac{187}{3840} \right)
\nonumber \\
&& + \left( 1 - \frac{m}{M} \right)^{6} 
\left( \frac{135311}{352800\pi^2} + \frac{1}{64}\ln 2 - \frac{13439}{322560} \right)
\nonumber \\ 
&& + \left( 1 - \frac{m}{M} \right)^{7} 
\left( \frac{39721}{117600\pi^2} + \frac{1}{128}\ln 2 - \frac{1427}{40320} \right)
\label{avg-y}
\\
&& + \left. \left( 1 - \frac{m}{M} \right)^{8} 
\left( \frac{5683891}{19051200\pi^2} + \frac{1}{256}\ln 2 - \frac{16901}{552960} \right) 
\right\} \ . \nonumber 
\ea

\begin{figure}[htp]
\psfig{figure=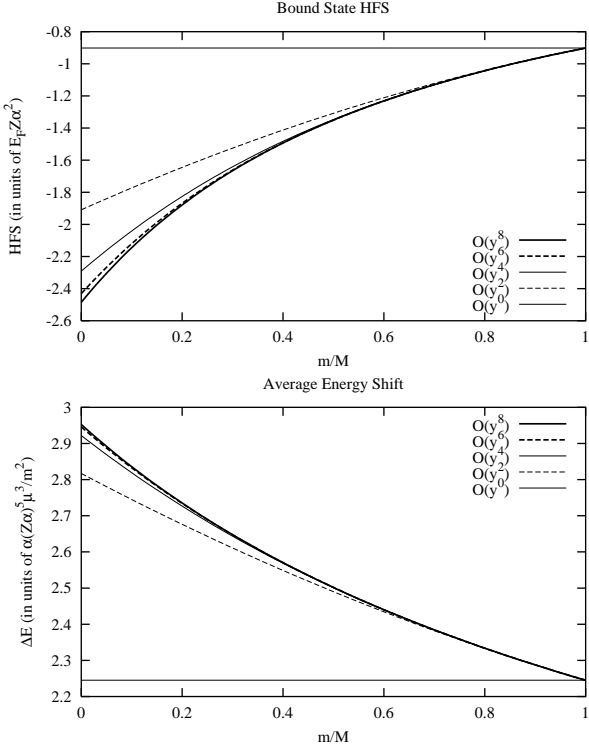,width=90mm}
\vspace*{3mm}
\caption{${\mathcal{O}}(\alpha(Z\alpha)^5)$ radiative recoil contributions to the hyperfine splitting and average energy shift in the $1-m/M$ expansion.}
\label{y-fig}
\end{figure}

In Fig.~\ref{y-fig} we illustrate the convergence of these series.  Both graphs suggest that the terms we have calculated in this expansion should yield reliable results for $m/M$ larger than about 0.15.   
\vspace{-10mm}

\begin{figure}[htp]
\psfig{figure=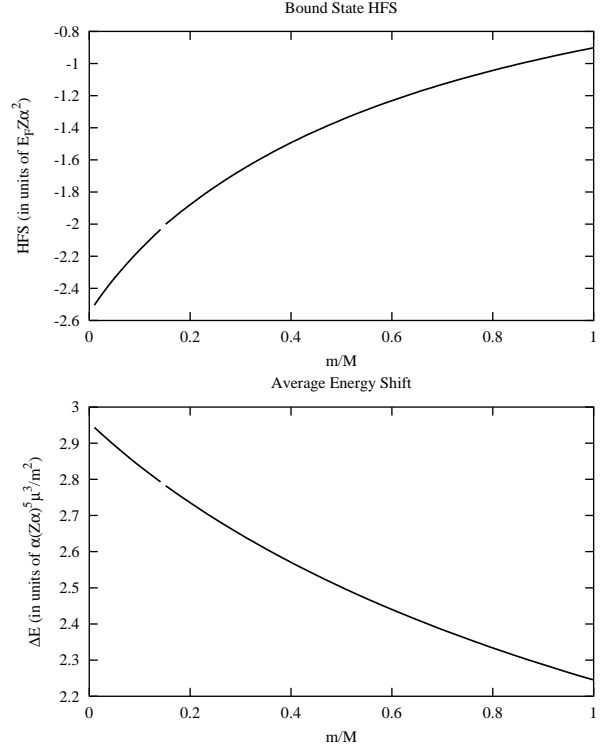,width=90mm}
\vspace*{3mm}
\caption{${\mathcal{O}}(\alpha(Z\alpha)^5)$ radiative recoil contributions to the hyperfine splitting and average energy shift for arbitrary values of $m$ and $M$.}
\label{splice-fig}
\end{figure}

Although neither expansion can handle arbitrary values of $m$ and $M$, the bound-state energy level corrections for any $m/M$ ratio can be reliably calculated with one of the expansions.  To illustrate this, we have spliced together the two expansions in Fig.~\ref{splice-fig}.  The expansions merge nicely at an $m/M$ value of 0.15, thereby providing a useful check on these methods.  It is also important, in the context of future applications of these methods, to have covered the entire range of $m/M$.

We have also calculated the corresponding ${\cal O}((Z\alpha)^6)$ pure recoil corrections.
For the hyperfine splitting we obtain
\ba
&& \delta E_{\rm hfs}^{\rm rec\ hard} \simeq \frac {8 (Z\alpha)^6 \mu^3}{3mM} 
\left\{ \frac {m}{M} \left ( -\frac{1}{2\ep} -6\ln 2 + \frac {5}{12} \right) \right. \nonumber \\
&& 
+\left({m\over M}\right)^2 \left ( \frac{1}{\ep} + \frac{9}{2\pi^2}\ln^2{M\over m} + \frac{27}{2\pi^2}\ln{M\over m} - \ln{M\over m} \right.
\nonumber \\ 
&&\qquad  \qquad  \left. - \frac {23}{12} + \frac{33\zeta_3}{\pi^2} + \frac{93}{4\pi^2} \right) \nonumber \\
&&
\left. + \left( {m\over M}\right)^3 \left( -\frac{3}{2\ep} - 6\ln 2 + \frac{25}{12} \right) \right\} . 
\label{hfs-rec}  
\ea
The divergences in this result are canceled by soft-scale terms, which can be calculated by extending the calculation of~\cite{Czarnecki:1998zv,Czarnecki:1999mw} to the unequal mass case, resulting in
\ba
&& \delta E^{\rm rec\ soft}_{\rm hfs} = \frac{(Z\alpha)^6 \mu^5}{m^2 M^2} \nonumber \\
&& \times \left[ -\frac{16}{3} \left (\log(2 \mu \alpha)  -\frac{1}{4\epsilon}  \right )
 + 4 \frac{mM}{\mu^2} + \frac{230}{27} \right] \ .
\label{hfs-rec-soft}
\ea
Combining the results of Eqs.~(\ref{hfs-rec}) and~(\ref{hfs-rec-soft}), we find that the total $(Z\alpha)^6$ pure recoil contribution to the hyperfine splitting is
\ba
\delta E^{\rm rec}_{\rm hfs} & \simeq & \frac {8 (Z\alpha)^6 \mu^3}{3mM} \left\{ \frac{3}{2} + \frac{\mu^2}{mM} \left[ \frac{65}{18} - 8\ln 2 + 2\ln (Z\alpha)^{-1} \right. \right. \nonumber \\
& & + \left( \frac{m}{M} \right) \left( \frac{9}{2\pi^2}\ln^2{M\over m} + \frac{27}{2\pi^2}\ln{M\over m} - \ln{M\over m} \right. \nonumber \\
& & \qquad \qquad \left. - 12\ln 2 - \frac{13}{12} + \frac{33\zeta_3}{\pi^2} + \frac{93}{4\pi^2} \right) \nonumber \\
& & + \left( \frac{m}{M} \right)^2 \left( \frac{9}{\pi^2}\ln^2{M\over m} + \frac{27}{\pi^2}\ln{M\over m} - 2\ln{M\over m} \right. \nonumber \\
& & \qquad \qquad \left. \left. \left. -13\ln 2 - \frac{4}{3} + \frac{66\zeta_3}{\pi^2} + \frac{93}{2\pi^2} \right) \right] \right\} \ .
\label{hfs-rec-tot}
\ea
The terms in the first line of Eq.~(\ref{hfs-rec-tot}) are in agreement with the result first obtained in~\cite{Bodwin:1982dx}.  The remaining terms, arising solely from the hard-scale contributions in Eq.~(\ref{hfs-rec}), can be used to obtain an analytic approximation to the function $f(x)$ near $x=1$ in Eq.~(72) of~\cite{KPfx}. 

For the hard contribution to the spin-independent energy shift we find
\ba
&& \delta E_{\rm aver}^{\rm rec\ hard} \simeq (Z\alpha)^6
 \frac {\mu^3}{m^2} \left \{ \frac {m}{M} \left( 4\ln 2 -\frac{7}{2} \right)
\right.  \nonumber \\
&& 
+ \left({m\over M}\right)^2 \left( \frac{4}{\pi^2}\ln{M\over m} - \frac{8}{3}\ln{M\over m} -\frac{12\zeta_3}{\pi^2} + \frac{3}{\pi^2} + \frac{8}{3} \right) \nonumber \\ 
&&
+\left({m\over M}\right)^3 \left ( 4\ln 2 - \frac{31}{6} \right) \nonumber \\
&&
+ \left({m\over M}\right)^4 \left(  -\frac{11}{3\pi^2}\ln^2{M\over m} - \frac{113}{18\pi^2}\ln{M\over m} - 2\ln{M\over m} \right. \nonumber \\
&& \qquad \qquad \left. \left. - \frac{6\zeta_3}{\pi^2} - \frac{1565}{72\pi^2} + \frac{62}{9} \right) \right\} . 
\label{avg-rec} 
\ea
The $m/M$ term of this expansion is in agreement with a calculation in Ref.~\cite{PG95}.  To our knowledge, the subsequent terms of this expansion are new.  In addition, we have calculated these energy level shifts as an expansion in $(1-m/M)$, but for brevity we shall omit these results. 

In spite of the fact that the hard-scale contribution to the average energy shift given by Eq.~(\ref{avg-rec}) is finite at this order, soft contributions are also present and are needed to arrive at the physical result for this quantity.  These soft contributions can be obtained by a calculation completely analagous to the one that produced Eq.~(\ref{hfs-rec-soft}).

\section{Conclusions}
We have demonstrated a method by which the corrections to the energy levels of a QED bound-state, with constituents of mass $m$ and $M$, can be expanded in either powers of $m/M$ or $(1-m/M)$.  Both expansions are applied directly to the integrands of the loop integrals arising from the hard-scale contributions to the energy shifts.  We have demonstrated the utility of these procedures by computing several terms in the expansions for the $\alpha (Z\alpha)^5$ radiative recoil and $(Z\alpha)^6$ pure recoil corrections to both the average energy shift and the hyperfine splitting of a general QED bound-state.  

Further studies of QED bound-state problems, using the methods described in this paper, might involve higher-order corrections to the energy level shifts.  Even in the absence of a complete calculation of such terms, it might be feasible to extract the terms enhanced by one or more factors of $\ln (M/m)$ by examining the singularities of the contributions from different expansion regions.  Since these singularities must cancel in the complete result, their coefficients can be found by a partial calculation of the divergent parts of those contributions which can be evaluated most easily.    

In a more general context, these expansion techniques are applicable to a plethora of other types of problems involving multiloop calculations with more than one external scale.  Many kinds of superficially disparate physical problems often depend on a few common classes of loop integrals, thereby reducing the number of technical hurdles which restrict the progress of precision calculations in particle physics.

{\it Acknowledgments} 
This work was supported in part by the Natural Sciences and
Engineering Research Council of Canada and the DOE under grant number
DE-AC03-76SF00515.

\section*{Appendix A: Calculation of Eikonal Integrals}

An expansion in powers of $m/M$ typically gives rise to integrals containing
eikonal propagators such as $(2kp{\,\pm\,}i\delta)$, arising from expansions in the momentum region where the loop momenta are all $\sim m$.  The easiest way to solve them is to
employ the integration-by-parts techniques \cite{Tkachev:1981wb,che81} so that 
any integral of the form in Eqs.~(\ref{E1}) and~(\ref{E2}) 
can be algebraically
expressed as a combination of the  
two-loop on-shell self-energy integrals and 
four new master integrals. The latter
are the only integrals we have to compute, and the results read
\ba
\label{j1}
&& J_{1}^{\pm} = \int \frac {[{\rm d}^D k_1] [{\rm d}^D k_2]}
{(k_1 Q -1 \pm i \delta) (k_2^2+i\delta) 
\left[ (k_1+k_2)^2 -1 + i \delta\right] }
\nonumber \\
=&& {1\over (4\pi)^D} \left[
2\Gamma(1-\ep) \Gamma(3\ep-2) B(4\ep-3,2\ep-1) 
\right. \nonumber \\
&& \left. 
-(1 \mp 1)
 \sqrt{\pi}
\Gamma\left(2\ep-{3\over 2}\right) 
B\left(\frac {5}{2}-3\ep,-\frac {1}{2}+\ep\right)
 \right],
\ea
\ba
\label{j2}
J_{2}^{\pm} =&& \int \frac {[{\rm d}^D k_1] [{\rm d}^D k_2]}
{(k_1 Q \pm i \delta) (k_2^2 - 1 +i\delta) 
\left[ (k_1+k_2)^2 -1 + i \delta\right] }
\nonumber \\
&&=
\pm \frac {\sqrt{\pi}}{(4\pi)^D}
\Gamma\left(2\ep-\frac {3}{2}\right) 
B\left(-\frac {1}{2}+\ep ,-\frac {1}{2}+\ep\right).
\ea
Please note that there is a typographical error for $J_{2}^{\pm}$ in \cite{Czarnecki:2000fv}.  For clarity, we now outline the process by which these master integrals are calculated.

Starting with $J_{1}^{+}$, we can Wick rotate the momenta and subsequently ignore the poles.  After shifting $k_{1}$ to $k_{1} - k_{2}$ and using the identity
\be
\label{georgi}
\frac{1}{A^{\alpha}B^{\beta}} = \frac{1}{B(\alpha,\beta)} \int_{0}^{\infty} d\lambda \frac{\lambda^{\beta - 1}}{[A + B\lambda]^{\alpha + \beta}} \ \ ,
\ee
we have
\ba
J_{1}^{+} & = & \int \frac{[d^{D}k_{1}][d^{D}k_{2}]}{k_{2}^{2}}  \nonumber \\
&& \times \int_{0}^{\infty} \frac{d\lambda}{[k_{1}^{2} + 1 + \lambda(k_{1}Q - k_{2}Q + 1)]^2} \ \ .
\ea
The $k_{1}$ integral can be evaluated, after completing the square and using $Q^{2}=-1$, so that
\be
J_{1}^{+} = \frac{\Gamma(\ep)}{(4\pi)^{D/2}} \int_{0}^{\infty} d\lambda \int \frac{[d^{D}k_{2}]}{k_{2}^{2} \left( \frac{\lambda^{2}}{4} + \lambda + 1 - \lambda k_{2}Q \right)^{\ep}} \ \ . 
\ee
Introducing a second parameter from the identity in Eq.~(\ref{georgi}), the $k_{2}$ integral can be evaluated, yielding
\ba
J_{1}^{+} & = & \frac{\Gamma(2\ep-1)}{(4\pi)^{D}} \int_{0}^{\infty} d\lambda \nonumber \\
&& \times \int_{0}^{\infty} d\rho \ \rho^{-\ep} \left( \frac{\rho \lambda^{2}}{4} + \frac{\lambda^{2}}{4} + \lambda + 1 \right)^{1-2\ep}
\ea
With the substitution
\be
\rho = 4z \ \frac{(\lambda/2+1)^2}{\lambda^{2}} \ \ ,
\ee
we can readily integrate over the remaining parameters to obtain the result for $J_{1}^{+}$ in Eq.~(\ref{j1}).  The evaluation of $J_{2}^{+}$ starts by introducing a Feynman parameter to combine and integrate over the $k_{2}$-dependent factors, and is followed by the introduction of a parameter from the identity in Eq.~(\ref{georgi}).

Turning to $J_{1}^{-}$, we note that the $-i\delta$ pole term in the eikonal propagator prohibits a Wick rotation.  Instead, the identity
\be
\int_{-\infty}^{\infty} \frac{dx}{x-a\pm i\delta} = P \int_{-\infty}^{\infty} \frac{dx}{x-a} \mp i\pi \delta(x-a) \ \ ,
\ee
with respect to the $k_{1}^{0}$ integral, allows us to write
\be
J_{1}^{-} = J_{1}^{+} + \Delta J_{1}
\ee
with
\be
\Delta J_{1} = 2\pi i \int \frac{[d^{D}k_{1}][d^{D}k_{2}] \ \delta(k_{1}^{0} - 1)}{(k_{2}^{2} + i\delta) [(k_{1}+k_{2})^{2} - 1 + i\delta] } \ \ .
\ee
A Feynman parameter can be introduced to evaluate the $k_{2}$ integral in $\Delta J_{1}$.  After performing the $k_{1}^{0}$ integral, we have
\ba
\Delta J_{1} & = & \frac{-2\pi \Gamma(\ep)}{(4\pi)^{D/2}(2\pi)^{D}} \int_{0}^{1} dx \nonumber \\
&& \times \int \frac{d^{D-1}{\bf{k}}}{\left[ x-x(1-x)+x(1-x){\bf{k}}^2 \right]^{\ep}} \ \ ,
\ea
where $\bf{k}$ denotes the $(D-1)$-dimensional spatial momentum associated with $k_{1}$.  Working in hyperspherical coordinates, and using
\be
\int d^{D-1}\Omega = \frac{2 \pi^{(D-1)/2}}{\Gamma \left( \frac{D-1}{2} \right)} \ \ ,
\ee
we obtain
\ba
\Delta J_{1} & = & -\frac{4\sqrt{\pi}}{(4\pi)^{D}} \frac{\Gamma(\ep)}{\Gamma \left( \frac{3}{2} - \ep \right) } \int_{0}^{1} dx \nonumber \\
&& \times \int_{0}^{\infty} \frac{k^{D-2} \ dk}{\left[ x^2 + x(1-x)k^{2} \right]^{\ep}} \ \ ,
\ea
with $k = |{\bf{k}}|$.  With the substitution
\be
k = \sqrt{\frac{x}{1-x}} \ z \ \ ,
\ee
the $x$ and $z$ integrals can be evaluated in terms of $B$-functions, yielding the result in Eq.~(\ref{j1}).  The evaluation of $J_{2}^{-}$ proceeds in a similar fashion.


\end{document}